\documentclass[12pt]{article}
\pdfoutput=1

\usepackage{cite}
\usepackage{booktabs}
\usepackage[english]{babel}
\usepackage{amsmath,amssymb,amsbsy,amstext, amsthm, simplewick}
\usepackage{hyperref}
\usepackage{graphicx}
\usepackage{amsfonts}
\usepackage{amssymb}
\usepackage[small]{caption}
\usepackage{upgreek}
\usepackage[svgnames,dvipsnames,x11names,table]{xcolor}
\usepackage{multirow}
\usepackage{geometry}
\usepackage[hang,flushmargin]{footmisc}
\usepackage{bm}
\usepackage{braket}
\usepackage{subcaption}
\usepackage{mathtools}
\usepackage{setspace}
\usepackage{cleveref}
\usepackage{comment}
\usepackage{scalerel}
\usepackage[normalem]{ulem}
\usepackage{slashed}
\usepackage{enumitem}
\usepackage{dsfont}
\usepackage{tikz}
\usepackage{colortbl}
\usepackage[table]{xcolor}
\usetikzlibrary{decorations.markings}
\usetikzlibrary{shapes.misc}

\graphicspath{{./figures/}}

\definecolor{colorC0}{HTML}{1f77b4}
\definecolor{colorC1}{HTML}{ff7f0e}
\definecolor{colorC2}{HTML}{2ca02c}
\definecolor{colorC3}{HTML}{d62728}
\definecolor{colorC4}{HTML}{9467bd}
\definecolor{colorC5}{HTML}{8c564b}

\newcolumntype{d}{>{\columncolor{colorC0!10}}c}
\newcolumntype{e}{>{\columncolor{colorC1!10}}c}
\newcolumntype{f}{>{\columncolor{colorC2!10}}c}
\newcolumntype{g}{>{\columncolor{colorC3!10}}c}

\makeatletter
\g@addto@macro\bfseries{\boldmath}
\makeatother

\hypersetup{
    colorlinks=true,
    linkcolor={red!50!black},
    citecolor={blue!50!black},
    urlcolor={blue!80!black}
}

\usepackage{colortbl}

\makeatletter
\newlength{\apb@width}
\newcommand{\autoparbox}[2][c]{\settowidth{\apb@width}{#2}\parbox[#1]{\apb@width}{#2}}

\makeatother

\definecolor{lightgray}{gray}{0.9}

\usepackage[framemethod=default]{mdframed}
\newmdenv[skipabove=7pt,
skipbelow=7pt,
rightline=false,
leftline=false,
topline=false,
bottomline=false,
backgroundcolor=gray!10,
linecolor=gray,
innerleftmargin=5pt,
innerrightmargin=5pt,
innertopmargin=5pt,
innerbottommargin=5pt,
leftmargin=0cm,
rightmargin=0cm,
linewidth=4pt]{eBox}

\usepackage[most]{tcolorbox}
\tcbset{colback=white, colframe=black,
        highlight math style= {enhanced, 
            colframe=red,colback=red!10!white,boxsep=0pt}
        }
\definecolor{light-gray}{gray}{0.95}

\crefname{table}{Table}{Tables}
\crefname{equation}{Eq.}{Eqs.}
\crefname{appendix}{App.}{Apps.}
\crefname{section}{Sec.}{Secs.}
\crefname{figure}{Fig.}{Figs.}

\numberwithin{equation}{section}

\def\beq{\begin{equation}}
\def\eeq{\end{equation}}

\def\bea{\begin{eqnarray}}
\def\eea{\end{eqnarray}}

\def\beq{\begin{equation}}
\def\eeq{\end{equation}}
\def\bea{\begin{eqnarray}}
\def\eea{\end{eqnarray}}

\def\Mpl{M_{\rm pl}}

\DeclareRobustCommand{\SkipTocEntry}[4]{}

\definecolor{colorTC}{rgb}{.2,.7,.2}

\definecolor{amethyst}{rgb}{0.6, 0.4, 0.8}

\definecolor{acolor}{rgb}{0.4, 0.2, 0.4}

\definecolor{blue3}{RGB}{31, 119, 180}
\definecolor{red3}{RGB}{	214, 39, 40}
\definecolor{orange3}{RGB}{255, 127, 14}
\definecolor{green3}{RGB}{44, 160, 44}

\begin{document}

\begin{titlepage}
\setcounter{page}{1} \baselineskip=15.5pt
\thispagestyle{empty}
$\quad$
\vskip 70 pt

\begin{center}

{\fontsize{20.74}{24} \bf The Cosmological
Preference \\[5pt] for Negative Neutrino Mass}
\end{center}

\vskip 20pt
\begin{center}
\noindent
{\fontsize{12}{18}\selectfont Daniel Green$^1$ and Joel Meyers$^2$}
\end{center}

\begin{center}
\vskip 4pt
\textit{$^1${\small Department of Physics, University of California, San Diego,  La Jolla, CA 92093, USA}\\
$^2${\small Department of Physics, Southern Methodist University, Dallas, TX 75275, USA}\\
}
\end{center}

\vspace{0.4cm}
 \begin{center}{\bf Abstract}
 \end{center}

\noindent
The most precise determination of the sum of neutrino masses from cosmological data, derived from analysis of the cosmic microwave background (CMB) and baryon acoustic acoustic oscillations (BAO) from the Dark Energy Spectroscopic Instrument (DESI), favors a value below the minimum inferred from neutrino flavor oscillation experiments.  We explore which data is most responsible of this puzzling aspect of the current constraints on neutrino mass and whether it is related to other anomalies in cosmology. We demonstrate conclusively that the preference for negative neutrino masses is a consequence of larger than expected lensing of the CMB in both the two- and four-point lensing statistics. Furthermore, we show that this preference is robust to changes in likelihoods of the BAO and CMB optical depth analyses given the available data. We then show that this excess clustering is not easily explained by changes to the expansion history and is likely distinct from the preference for for dynamical dark energy in DESI BAO data. Finally, we discuss how future data may impact these results, including an analysis of Planck CMB with mock DESI 5-year data. We conclude that the negative neutrino mass preference is likely to persist even as more cosmological data is collected in the near future.

\end{titlepage}
\setcounter{page}{2}

\restoregeometry

\begin{spacing}{1.2}
\newpage
\setcounter{tocdepth}{2}
\tableofcontents
\end{spacing}

\setstretch{1.1}
\newpage

\section{Introduction}

The first year baryon acoustic oscillation (BAO) results from the Dark Energy Spectroscopic Instrument (DESI)~\cite{DESI:2024uvr,DESI:2024mwx}, when combined with data from the cosmic microwave background (CMB)~\cite{Planck:2018vyg,Planck:2018lbu,ACT:2023dou}, gives a surprisingly strong constraint on the sum of neutrino masses. The 95\% upper limit, $\sum m_\nu < 70$ meV~\cite{DESI:2024mwx} is significantly stronger than one would anticipate from Fisher forecasts~\cite{Font-Ribera:2013rwa,CMB-S4:2016ple,SimonsObservatory:2018koc}. It was shown in Ref.~\cite{Craig:2024tky} that this can be understood as a preference for negative neutrino masses. The energy density carried by the (physical) cosmic neutrino background is positive and gives a definition of the mass that is similarly positive, $\sum m_\nu \geq 0$~\cite{Lesgourgues:2006nd,TopicalConvenersKNAbazajianJECarlstromATLee:2013bxd,Gerbino:2018jee,Dvorkin:2019jgs}. When the cosmological definition of neutrino mass is extended to allow $\sum m_\nu<0$, the resulting analysis yields $\sum m_\nu = -160 \pm 90$ meV~\cite{Craig:2024tky} which excludes at $3\sigma$ even the minimum sum of masses inferred from neutrino flavor oscillation experiments~\cite{ParticleDataGroup:2020ssz}. 

Other tensions are known to exist when combining data from multiple surveys~\cite{Abdalla:2022yfr}. The most well known cosmological tension is the Hubble tension between the CMB+BAO measurement of $H_0$~\cite{Planck:2018vyg} and the distance ladder measurement~\cite{Riess:2021jrx}. However, even Planck and DESI BAO alone show a modest statistical preferences for dynamical dark energy~\cite{DESI:2024mwx,DESI:2024kob}. Although the precise origin and meaning of this preference is unclear~\cite{Cortes:2024lgw,Colgain:2024xqj,Colgain:2024ksa,Carloni:2024zpl,Wang:2024rjd}, one would still like to understand to what degree the behavior of $\sum m_\nu$ is just a symptom of the known tensions in the expansion history. Neutrino mass leaves a unique signal in the clustering of matter and it is therefore plausible that this feature of the current cosmological data is distinct from other anomalies.

The conventional method to achieve a measurement of neutrino mass with cosmological observations is a measurement of the suppression of clustering due to free-streaming neutrinos~\cite{Lesgourgues:2006nd,Green:2021xzn}. While the neutrinos also directly affect the rate of expansion and growth of structure at late times, given existing constraints, neutrinos make up only a sub-percent fraction of the matter density in the universe, and the imprints of neutrino mass through these effects are too small to be measurable with current data~\cite{Yu:2018tem,Brinckmann:2018owf,Sailer:2021yzm}. Because matter clustering is an integrated effect that accumulates throughout cosmic history, the impact of the neutrino mass on matter clustering is significantly enhanced to at least the few percent level.

The unfortunate consequence of the nature of this measurement is that it relies on knowledge of the optical depth to reionization and the amplitude of CMB lensing~\cite{Allison:2015qca}, in addition to the measurement of the expansion history. As a result, even with improvements in BAO measurements with future data from DESI, it is unclear if we are likely to get more clarity on the nature of neutrinos for the foreseeable future. It is therefore essential to understand exactly what is driving the current neutrino mass measurement towards negative values and identify data that could clarify the situation without waiting for a future CMB measurement of the optical depth from the ground~\cite{Essinger-Hileman:2014pja}, balloon~\cite{Errard:2022fcm}, or satellite~\cite{LiteBIRD:2022cnt}. This is not only important for the measurement of neutrino mass itself but also impacts how we think of the relationship between cosmic surveys and lab-based neutrino experiments~\cite{Gerbino:2022nvz}.

In this paper, we will break down the measurement of $\sum m_\nu$, and its generalization to negative mass, to identify the data and physical effects that are most relevant to the signal. First, we will show that the current measurement is driven by the four-point lensing information in the CMB, while being consistent with the two-point lensing. This conclusion is not altered by changes to the Planck likelihood (optical depth) or the specific BAO data (matter density). In this sense, the inferred value of the neutrino mass points unambiguously towards an excess of clustering.

We will then discuss the origin of excess clustering by comparing the impact of dynamical dark energy with negative neutrino mass. Changing the expansion history impacts the growth function and thus the amplitude of the matter power spectrum. As a result, it is tempting to believe the preference of negative neutrino mass is just another indication of a more complicated expansion history. However, any change to the expansion history must also be consistent with measurements via the BAO. Furthermore, because the CMB lensing signal is generally sensitive to higher redshifts than the era of dark energy domination, large changes to the equation of state of the dark energy have much more muted impact on the lensing potential than on the expansion history. Indeed, we will find that analyses allowing for both dynamical dark energy and negative neutrino mass provide posteriors that remain peaked at negative neutrino mass values. In this sense, we do not find evidence that apparent anomalies in the expansion history are related to the larger than expected CMB lensing amplitude.

Finally, we will explore how future data may alter this conclusion or inform our understanding the physical origin of this expected signal. We will explore the impact on future DESI data by creating mock 5-year DESI BAO data designed to be compatible with Planck TTTEEE best-fit $\Lambda$CDM cosmology.  We show that CMB data combined with this mock BAO data requires a large upward shift in the BAO data to be compatible with positive neutrino mass. Therefore, without a major systematic error affecting the existing DESI data, the CMB lensing amplitude is likely to remain larger than would be expected in a model containing the minimum sum of neutrino masses.

This paper is organized as follows: In Section~\ref{sec:lensing}, we discuss the origin of the preference for negative neutrino mass with current data. In Section~\ref{sec:darkenergy}, we show that the preference for negative neutrino mass is not resolved by changing the expansion history at late times. Finally, in Section~\ref{sec:future}, we discuss the potential impact of future data. We conclude in Section~\ref{sec:conclusion}.

Throughout the paper, we will refer to both the physical neutrino mass and its extension to negative values as $\sum m_\nu$ whenever the distinction is either unimportant or clear from context.  When the difference is important, we refer to the negative mass extension as $\sum \tilde{m}_\nu$ and use $\sum m_\nu$ for the physical neutrino mass subject to the constraint $\sum m_\nu \geq 0$.

\section{The Origin of Negative Neutrino Mass}\label{sec:lensing}

In this section, we will demonstrate that the preference for negative neutrino mass is driven by a measurement of the CMB lensing amplitude that is larger than expected. First, we will explain the physical reason that CMB lensing is the main source of sensitivity to $\sum m_\nu$ and then explore how constraints on $\sum m_\nu$ are influenced by the inclusion of different types of data.

Throughout this paper, we use the same prescription for negative neutrino mass as described in Ref.~\cite{Craig:2024tky}.  Specifically, the effects of neutrino mass are modeled by through a parameter $\sum \tilde{m}_\nu$ which imposes a rescaling of the CMB lensing potential power spectrum that is designed to match the effect of physical neutrino mass in the $\sum m_\nu \geq 0$ regime and which leads to an enhancement of the lensing power spectrum in the $\sum \tilde{m}_\nu < 0$ regime.  The effect of $\sum \tilde{m}_\nu$ is consistently applied to $C_L^{\phi\phi}$ and the lensed $C_\ell^{TT}$, $C_\ell^{TE}$, and $C_\ell^{EE}$ spectra.    We implement this parameterized neutrino mass in a modified version of \texttt{CAMB}~\cite{Lewis:1999bs,Howlett:2012mh}, which we use for all of our Boltzmann calculations.  Unless otherwise stated, we employ the likelihood for CMB temperature and polarization from Planck's 2018 data release~\cite{Aghanim:2019ame}, along with the combination of ACT DR6~\cite{ACT:2023dou,ACT:2023kun} and Planck CMB lensing~\cite{Carron:2022eyg}, and DESI BAO~\cite{DESI:2024lzq,DESI:2024mwx,DESI:2024uvr}.  This combination of data matches what was used by the DESI team to derive cosmological constraints~\cite{DESI:2024mwx}.  Our parameter constraints were obtained with \texttt{cobaya}~\cite{Torrado:2020dgo}, using the Markov chain Monte Carlo sampler adapted from \texttt{CosmoMC}~\cite{Lewis:2002ah,Lewis:2013hha} using the fast-dragging procedure~\cite{Neal:2005}.  We ran all chains until the Gelman-Rubin statistic was $R-1<0.01$. Results were analyzed and plotted using \texttt{GetDist}~\cite{Lewis:2019xzd}.

\subsection{Power of CMB Lensing}

Neutrino mass is measured through the gravitational influence of neutrinos on the visible universe through the homogeneous expansion, the growth of structure, the statistics of inhomogeneities, and gravitational lensing. Being gravitational in origin, all of these effects are determined by the total energy density in (non-relativistic) neutrinos
\beq
\Omega_\nu h^2=6 \times 10^{-4}\left(\frac{\sum m_\nu}{58  \, \mathrm{meV}}\right) \qquad f_\nu \equiv \frac{\Omega_\nu}{\Omega_m} = 4\times 10^{-3} \left(\frac{\sum m_\nu}{58 \, \mathrm{meV}}\right)   \ .
\eeq
Neutrinos with mass around 50~meV become non-relativisitc around $z_\nu \approx 100$.  Since the neutrinos only redshift like matter when they are non-relativists,  $\Omega_m (z > z_\nu) = \Omega_{\rm cdm} + \Omega_{\rm b}$ while $\Omega_m (z \ll z_\nu) = \Omega_{\rm cdm} + \Omega_{\rm b} + \Omega_\nu$. In principle, it is possible to measure $\sum m_\nu$ from comparing $\Omega_m(z \approx 1100)$ determined by the primary CMB and $\Omega_m(z ={\cal O}(1))$ from the BAO. Unfortunately, to place a competitive constraint via this comparison, we would need to measure $\Omega_m$ at 0.2\% precision at both early and late times. However, we note that Planck TTTEEE~\cite{Planck:2018vyg} determined $\Omega_m = 0.3166 \pm 0.0084$ (amounting to 2.7\% precision) and adding DESI BAO~\cite{DESI:2024mwx} yields $\Omega_{m}=0.307 \pm 0.005$ (1.6\% precision). As a result, current direct measurements of the expansion history alone do not meaningfully constrain $\sum m_\nu$.

The measurement of $f_\nu$ from $\Omega_m(z)$ is a concrete example of the broader challenge in measuring neutrino mass. In principle, massive neutrinos have a wide range of measurable effects on cosmic observables~\cite{Zhu:2013tma,Zhu:2014qma,Chiang:2017vuk,Giusarma:2018jei,Zhu:2019kzb,Shiveshwarkar:2023xjv,Nascimento:2023ezc}. However, for any measurement that is related to the instantaneous gravitational force, like the expansion or the growth rate, the effects on observables will be at the sub-percent level for $\sum m_\nu \sim 58$ meV and therefore we expect an uncertainty of $\sigma(\sum m_\nu) = {\cal O}(100 \, {\rm meV})$ for any percent-level measurement of these quantities.

On the other hand, cosmological clustering of matter is sensitive to the integrated expansion history and therefore is the exception to this general expectation. Because the neutrinos inhibit the growth of structure starting at $z_\nu \approx 100$, the matter power spectrum at late times, $z \ll z_\nu$, takes the form 
\beq
P^{\left(\sum m_\nu\right)}\left(k \gg k_{\mathrm{fs}}, z\right) \approx\left(1-2 f_\nu-\frac{6}{5} f_\nu \log \frac{1+z_\nu}{1+z}\right) P^{\left(\sum m_\nu=0\right)}\left(k \gg k_{\mathrm{fs}}, z\right) \, .
\eeq
The large logarithm from the integrated effect of expansion implies the suppression of power is enhanced by a factor of approximately $2+(6/5)\log 100 \approx 7.5 $. As a result, even a minimum sum of neutrino masses leaves a two percent suppression of power which is measurable with current precision, particularly by observing gravitational lensing of the CMB by the intervening matter~\cite{Kaplinghat:2003bh,ACT:2023dou,ACT:2023kun}.

Despite the above argument, the tendency for $\sum m_\nu < 0$ with current data is not necessarily due to a single measurement. First of all, both the CMB two-point and four-point statistics are sensitive to gravitational lensing~\cite{Lewis:2006fu} and it is possible that the preference for negative neutrino mass could be driven by only one of these two statistics. Furthermore, it is entirely possible that lensing is consistent with positive neutrino mass and the tendency for negative neutrino mass is just another symptom of peculiarities involving the inference of the late-time expansion. Given the existing tensions with $H_0$, and the hints in DESI for dynamical dark energy, it is important to verify the precise source of these unexpected results.

Even with this logarithmic enhancement, the measurement of $\sum m_\nu$ via CMB lensing requires a high-precision measurement of $\Omega_m$ from the expansion history, as measured by the BAO. Specifically, since the lensing power spectrum scales as $C_L^{\phi\phi} \sim \Omega_m^2$, the uncertainty on the log-enhanced neutrino suppression is still limited by $7.5 \sigma(f_\nu) \gtrsim 2 \sigma(\Omega_m)$, holding other parameters fixed. This behavior is shown to arise in detailed forecasts~\cite{Pan:2015bgi} and explains the improvement on neutrino mass constraints provided by DESI BAO. However, the limitations presented by the optical depth, $\tau$, imply that continued improvements in the measurement of $\Omega_m$ will have limited impact on $\sigma(\sum m_\nu)$ until $\tau$ is measured at higher precision~\cite{CMB-S4:2016ple,Abazajian:2019eic}.

\subsection{Two- versus Four-point Information}
\label{subsec:2pt4pt}

The CMB is gravitationally lensed, whether we care to look for the signal or not. Therefore, the lensing signal of neutrino mass impacts any basic summary statistic, including the TT, TE, and EE spectra. Of course, the lensing potential can be (optimally) reconstructed from the temperature and polarization maps~\cite{Hu:2001kj,Okamoto:2003zw}, and power spectrum of the reconstructed lensing potential can be calculated from particular configurations of the CMB trispectrum~\cite{Hu:2001fa}. As a result, CMB two-point and four-point statistics both carry information about the late universe, independent from the integrated expansion history.

Naturally, it is important to understand if the origin of the preference for negative neutrino mass appears in both the two- and four-point information. One significant effect of lensing on the the CMB power spectra (i.e.~two-point information) is to smooth the acoustic peaks~\cite{Lewis:2006fu}. In principle, a wide variety of modifications to the linear evolution could give rise to a similar suppression of the peaks. The preference for additional smoothing was observed in Planck data~\cite{Planck:2018vyg}, sometimes appearing as a larger value of $A_{\rm lens}$ for the power spectra~\cite{Calabrese:2008rt}. One might then like to determine whether the preference for $\sum m_\nu < 0$ is driven by the same smoothing effect alone.

The reconstruction of the CMB lensing potential is enabled by an entirely separate physical effect of lensing, the off-diagonal correlations of the temperature and polarization multipole moments $a_{\ell m}$ and $a_{\ell'  m'}$. The specific realization of the lensing potential breaks statistical isotropy of the observed temperature and polarization anisotropies and thus is encoded in these correlations. This cannot arise from changes to the linear evolution or other changes to the linear transfer functions associated with projection effects, etc. Given that the CMB fluctuations are observed to be very nearly Gaussian~\cite{Planck:2019kim}, introducing new sources of non-Gaussianity to explain the enhanced lensing amplitude is an equal, if not more, drastic alteration of the cosmological model as enhancing the late time clustering. In this sense, separating the effect of the CMB lensing power spectrum from the changes to the TT, TE, and EE power spectra is an important window into the origin of the preference for negative neutrino mass.

The challenge in isolating the four-point information is that the two-point lensing information is encoded in the temperature and polarization maps (and thus also the spectra) from the beginning. In principle, one can use delensing to remove this information, but this procedure requires much more precise measurements of the CMB than are currently available in order to be effective~\cite{Seljak:2003pn,Green:2016cjr,Hotinli:2021umk}. Instead, we will constrain $\sum m_\nu$ while we marginalize over the lensing amplitude only in the two-point statistics. We will refer to this parameter as $B_{\rm lens}$ to distinguish it from $A_{\rm lens}$ which would be applied to the lensing power spectrum as it appears in both two- and four-point statistics. We can also remove the four-point function entirely by dropping the $C_L^{\phi \phi}$ data from the analysis. 

\begin{figure}[t!]
    \centering
    \includegraphics[width=0.6\textwidth]{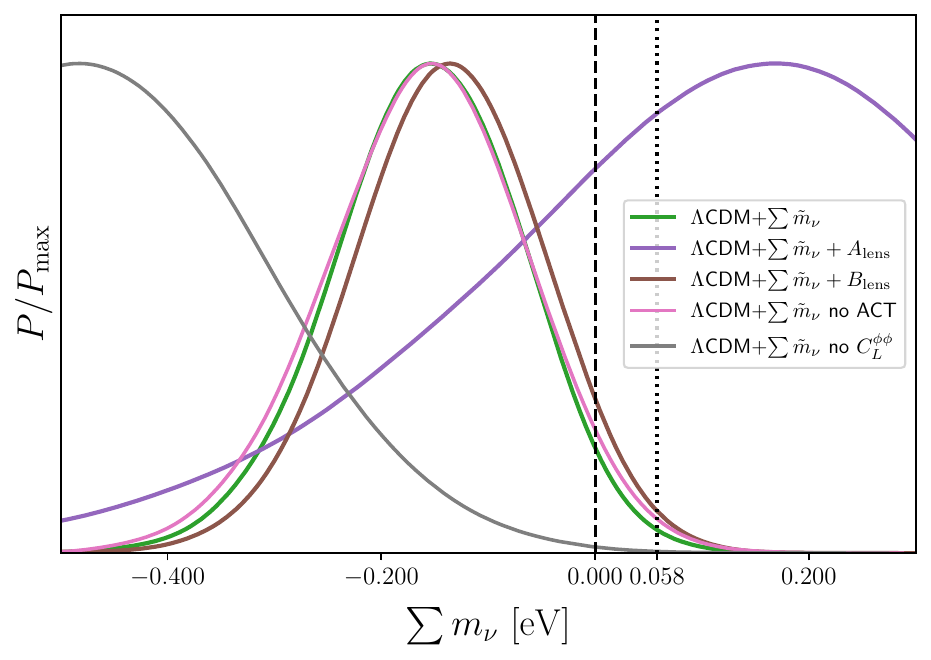}
    \caption{Posterior of neutrino mass in eV. The green line shows constraints from Planck + ACT lensing + DESI.  The purple line marginalizes also over $A_\mathrm{lens}$, a parameter that scales the CMB lensing power spectrum as $C_L^{\phi\phi} = A_\mathrm{lens} C_L^{\phi\phi,\mathrm{fid}}$ (which impacts both the lensed temperature and polarization power spectra and the reconstructed lensing power spectrum).  The brown line marginalizes instead over $B_\mathrm{lens}$, a parameter that scales the effect of lensing on the lensed temperature and polarization power spectra, but leaves the reconstructed lensing power spectrum unchanged. The pink line excludes the ACT lensing information.  The gray line does not use lensing reconstruction information from Planck or from ACT.}
    \label{fig:mnu_1d_lensing_variants}
\end{figure}

The results of our analysis are shown in Figure~\ref{fig:mnu_1d_lensing_variants}.  We see that the constraints on neutrino mass from the full analysis and the analysis with only four-point lensing information (marginalized over $B_{\rm lens}$) are nearly identical. In this precise sense, the preference for negative neutrino mass is not driven by the previous two-point lensing anomalies. The four-point lensing is measured with much higher signal to noise than the smoothing of the power spectrum~\cite{Planck:2018lbu,Carron:2022eyg,ACT:2023dou,ACT:2023kun}, and this analysis confirms that the four-point lensing information drives the constraints. The two-point only analysis (no $C_L^{\phi \phi}$) prefers even more negative values of $\sum m_\nu$, and thus is consistent with the full results but it is not the driving factor in the full result. Finally, to demonstrate that $A_{\rm lens}$ and $B_{\rm lens}$ are good proxies for the neutrino mass signal, we see that the constraints degrade dramatically and shift back to $\sum m_\nu > 0$ when we marginalize over $A_{\rm lens}$, which rescales both the two-point and four-point lensing amplitudes.

Naturally, one might speculate that a systematic error in Planck lensing reconstruction could account for the negative mass preference. We note that the inclusion of ACT lensing data does not impact the central value, suggesting that ACT and Planck lensing are consistent with each other. We will explore the potential for additional lensing data to inform our understanding of $\sum m_\nu$ in more detail in Section~\ref{sec:flens}.

\subsection{The Optical Depth and Planck Likelihoods}

A measurement of $\sum m_\nu$ from the suppression of power requires a precise and accurate measurement of the primordial amplitude of fluctuations, $A_s$. We measure this from the primary CMB, however, the high-$\ell$ CMB is mostly sensitive to $e^{-2\tau} A_s$ where $\tau$ is the optical depth to reionization. As a result, the determination of $\tau$ from the lower $\ell$ CMB (temperature and polarization) is an essential part of the neutrino mass measurement~\cite{Allison:2015qca}.

The central value of $\tau$ from the CMB has varied significantly over the years and is naturally a point of concern for the measurement of clustering in the late universe. However, as explained in Ref.~\cite{Craig:2024tky}, the large negative central value of $\sum \tilde m_\nu$ makes it implausible that the entire shift from the expected minimum value $\sum \tilde m_\nu = 58$ meV to $-160$ meV can be explained by a bias in the measurement of $\tau$. One illustration of the challenges presented by the optical depth measurement is the variety of $\tau$ values produced in different iterations of the Planck likelihood. The constraint on the optical depth reported in Planck 2018 results was $\tau = 0.054 \pm 0.007$~\cite{Planck:2018lbu}.  More recent analyses of Planck data with the SRoll2 map-making approach~\cite{Delouis:2019bub} have found larger values for the optical depth $\tau = 0.0627^{+0.0050}_{-0.0058}$~\cite{deBelsunce:2021mec} (see also Ref.~\cite{Pagano:2019tci}). 

\begin{figure}[ht!]
    \centering
    \includegraphics[width=\textwidth]{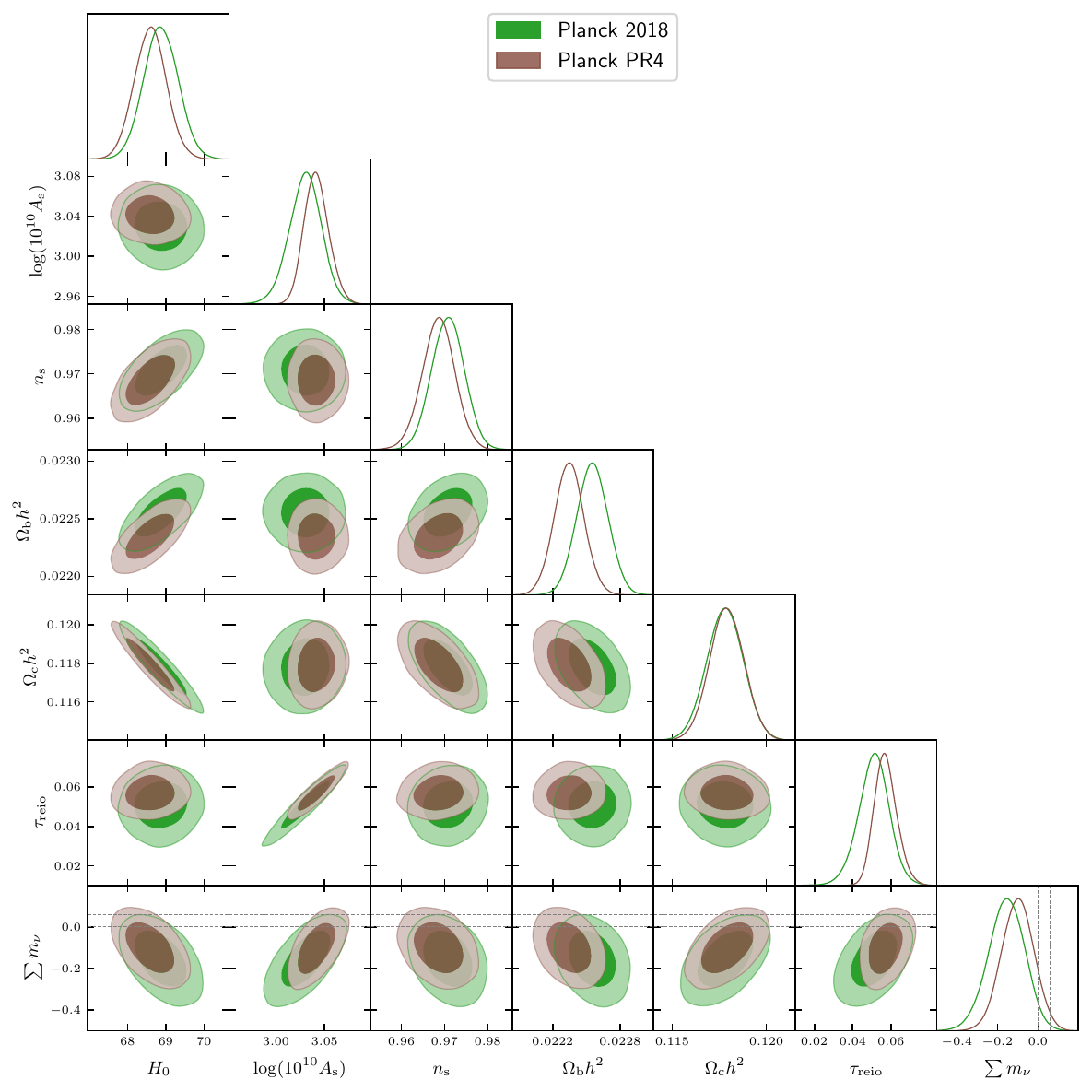}
    \caption{Triangle plot showing parameter constraints from Planck + ACT lensing + DESI BAO in the $\Lambda$CDM+$\sum \tilde{m}_\nu$ model for two different versions of the Planck likelihood.  The constraints labeled `Planck 2018' use the likelihoods that accompanied the Planck 2018 results~\cite{Planck:2019nip}, and those labeled `Planck PR4' use the NPIPE PR4 CamSpec high-$\ell$ likelihood~\cite{Rosenberg:2022sdy} along with the SRoll2 low-$\ell$ EE likelihood~\cite{Delouis:2019bub}.   Dashed lines show vanishing neutrino mass $\sum m_\nu = 0$ and the minimal sum of neutrino mass $\sum m_\nu = 58$~meV. Values of neutrino mass are reported in eV and $H_0$ in $\mathrm{km}\, \mathrm{s}^{-1}\, \mathrm{Mpc}^{-1}$.}
    \label{fig:triangle_PR4}
\end{figure}

In order to address the degree to which the preference for negative neutrino mass is affected by these differences, we present in Figure~\ref{fig:triangle_PR4} a comparison of the constraints derived from Planck 2018 and more recently released likelihoods.  One can see from these constraints that the preference for larger optical depth also leads to an upward shift in the posterior for neutrino mass, though negative neutrino mass is still favored in this case ($\sum \tilde{m}_\nu = -100 \pm 80$~meV). Importantly, the PR4 is also known to remove the preference for a large value of $A_{\rm lens}$ as measured by the two-point statistics~\cite{Tristram:2023haj}, but does not remove the preference for negative neutrino mass. This offers further confirmation that the four-point lensing is largely responsible, given that is the higher signal to noise measurement. Our results are consistent with a similar analysis of DESI and PR4 in~\cite{Allali:2024aiv}, although we emphasize that the shift towards positive neutrino mass in our analysis is not sufficient to shift the maximum of the posterior to positive values. 

\subsection{BAO with (e)BOSS and DESI Y1}

In addition to requiring a measurement of the optical depth in order to constrain $A_s$, observation of the BAO is required to measure $\Omega_m$ in order fix the amplitude of $C_L^{\phi \phi}$ that would be expected for $\sum m_\nu=0$. The lensing amplitude is determined by the total amount of non-relativistic matter and thus the enhancement compared to expectations in the total amount of lensing could be attributed to a bias in the value of $\Omega_m$ towards a smaller value.

One might naturally wonder if the measurement in $\Omega_m$ using DESI BAO could be sufficiently biased to explain the preference for negative neutrino mass. The value of $\Omega_m = 0.3069 \pm 0.0050$ (DESI+CMB)~\cite{DESI:2024mwx}  represents a downward shift from the Planck value of $\Omega_m = 0.3153 \pm 0.0073$~\cite{Planck:2018vyg}. Furthermore, several of the LRG data points with DESI appear to be in tension with measurements of the BAO at the same redshifts with (e)BOSS.

Although the preference for negative neutrino mass became more pronounced with DESI, the same feature was already noticed in the eBOSS final data release. Their measurement of $\sum m_\nu < 129 $~meV (95\% CL) using Planck + BAO~\cite{eBOSS:2020yzd} also represents a significantly more stringent upper limit than one might have expected. To explain this result, eBOSS fit their $\sum m_\nu > 0$ likelihood to a Gaussian and showed the peak of the Gaussian was at negative values of $\sum m_\nu$ (finding $\sum m_\nu = -26 \pm 74$ in their Gaussian fit). This was surely also the case in many other analysis such as that found upper limits much stronger than would be expected from Fisher forecasts, and thus require (by the Cramer-Rao bound) that the peak of a Gaussian likelihood would have to similarly be shifted to negative values as well. For example, the addition of Lyman~$\alpha$ forest or full shape information, the (e)BOSS+Planck bounds had already reached $\sum m_\nu < 90$ meV~\cite{Palanque-Delabrouille:2019iyz} and $\sum m_\nu < 82$ meV~\cite{Brieden:2022lsd} respectively. Like the DESI analysis, these results disfavor the inverted  hierarchy for neutrino mass and are significantly stronger than expectations from forecasts. Together, these results suggest that the preference for negative neutrino mass has been present in a number of analyses prior to DESI.

\begin{figure}[ht!]
    \centering
    \includegraphics[width=0.6\textwidth] {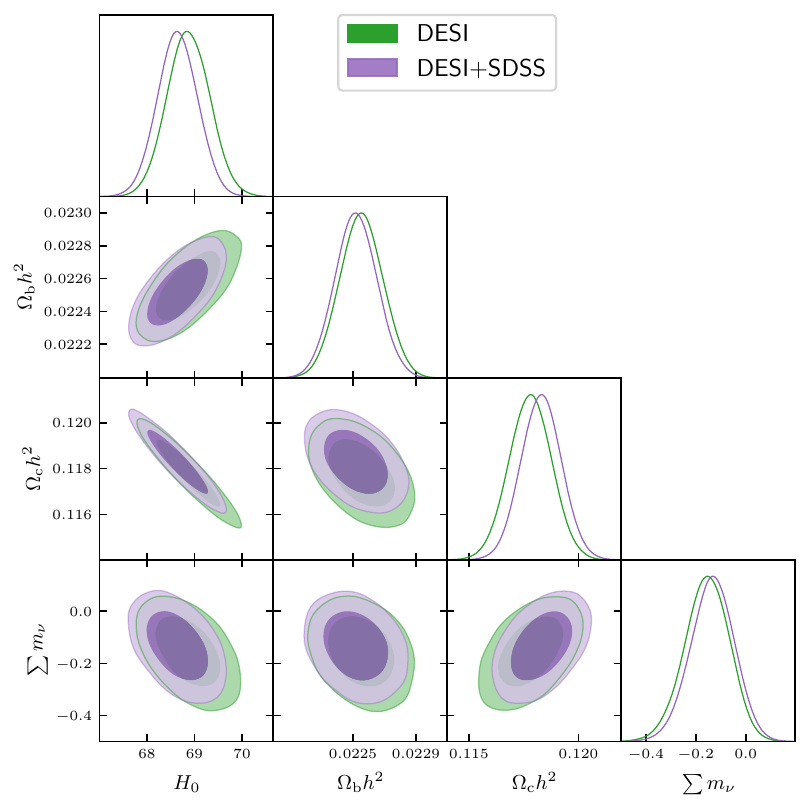}
    \caption{Triangle plot showing parameter constraints from Planck + ACT lensing + BAO data from either DESI or DESI + SDSS in the $\Lambda$CDM+$\sum \tilde{m}_\nu$ model.    Dashed lines show vanishing neutrino mass $\sum m_\nu = 0$ and the minimal sum of neutrino mass $\sum m_\nu = 58$~meV. Values of neutrino mass are reported in eV and $H_0$ in $\mathrm{km}\, \mathrm{s}^{-1}\, \mathrm{Mpc}^{-1}$.}
    \label{fig:triangle_BAO_variants}
\end{figure}

Of course, we can directly assess the impact of the discrepant points in the DESI BAO data by repeating the analysis using SDSS (BOSS and eBOSS) data in place of some of the DESI data. Here we follow the analysis performed in~\cite{DESI:2024mwx}: the SDSS data used in place of DESI  (BGS and LRG) for $z=0.15,0.38,0.51$ and combined DESI+SDSS for the Ly$\alpha$. The results of this analysis are shown in Figure~\ref{fig:triangle_BAO_variants}. We can see a small upward shift in $\sum m_\nu$ and $\Omega_c h^2$, that is expected from the slightly lower value of $\Omega_m$ between DESI and (e)BOSS, the shift does not meaningfully change the preference for negative masses. Moreover, the shift is sufficiently small that the conclusion is not easily altered by including more data either in addition to or in place of DESI.

\begin{figure}[ht!]
    \centering
    \includegraphics[width=0.6\textwidth] {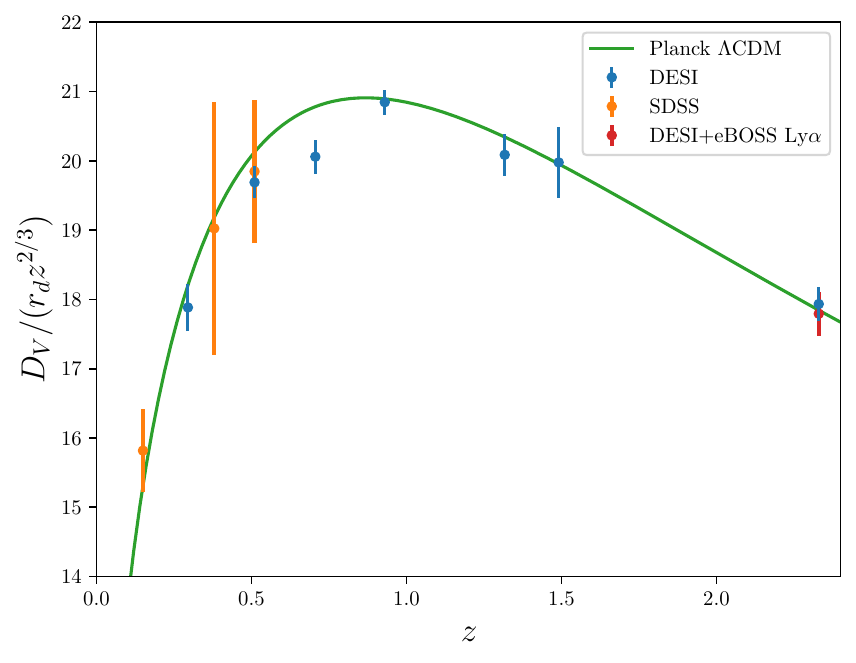}
    \caption{DESI and SDSS measurements of the angle-averaged distance $D_V=(zD_M^2D_H)^{1/3}$ divided by the radius of the sound horizon at the baryon drag epoch $r_d$, rescaled by a factor $z^{-2/3}$ to match the presentation in Ref.~\cite{DESI:2024mwx}.  The DESI+SDSS analysis drops the two lowest redshift bins from DESI (blue points) in favor of the three SDSS measurements shown here (orange points) and exchanges the highest redshift DESI measurement for the combined DESI+eBOSS Lyman-$\alpha$ measurement (red point).  Data for this plot came from Refs.~\cite{DESI:2024mwx} and \cite{eBOSS:2020yzd}.  Where measurements were reported in terms of $D_M$ and $D_H$ separately, those measurements were combined into $D_V$ for the purpose of this plot, though the likelihood utilized $D_M$ and $D_H$.}
    \label{fig:DV_over_rs_DESI_SDSS}
\end{figure}

\section{Expansion History and Clustering}\label{sec:darkenergy}

A central question about the tension between $\sum m_\nu \geq 58$ meV from neutrino oscillations and the current constraints from the CMB + BAO is whether it can be entirely attributed to unconventional dark energy or other anomalies in the measurement of the expansion history. While we generally expect constraints on $\sum m_\nu$ to degrade with a sufficiently flexible model of dark energy, that does not imply that it is the expansion itself that is responsible for the preference for negative neutrino mass. In this section, we will explore the relationship between expansion history and the growth of structure and how it impacts the neutrino mass constraints.

\subsection{Growth versus Expansion}

From our knowledge of the physical scales associated with acoustic peak locations of the CMB, the measurement of the BAO primarily determines the expansion history at low redshift. The physical distance to redshift $z$ is determined from the size of the BAO feature in angle and redshift. In $\Lambda$CDM, BAO data is most important for determining $\Omega_m$, as the density of radiation is too small to significantly impact the expansion history at $z<10$ and, therefore, $\Omega_\Lambda$ and $\Omega_m$ are related by the  flatness condition at late times ($\Omega_\Lambda+\Omega_m \approx 1$).

The amplitude of the (linear) matter power spectrum is determined by the primordial power spectrum, the growth function $D(z)$, and $\Omega_m H_0^2$ via
\beq
P_m(k) = \frac{4}{25} \frac{A_s D(z)^2}{\Omega_m^2 H_0^4} T(k)^2 \left(\frac{k}{k_\star}\right)^{n_s-1}
\eeq
where $k_\star$ is the pivot scale, $T(k)$ is the transfer function and $T(k \to 0) \to 1$.  Therefore in the limit $k \to 0$, the amplitude of $P_m(k)$ is determined by $A_s$, $D(z)$, and $\Omega_m H_0^2$. The growth function is determined by 
\beq
D(a) = \frac{5 \Omega_m}{2} \frac{H(a)}{H_0} \int_0^a \frac{d a^{\prime}}{\left(a^{\prime} H\left(a^{\prime}\right) / H_0\right)^3} \, ,
\eeq
where $D(z)\equiv D(a(z)=1/(1+z))$. From the Friedmann equation, we know at late times (where the radiation density is negligible)
\beq
H^2(a) = H_0^2\left( \Omega_m a^{-3} + \Omega_{\rm d.e.} a^{-3(1+w_0+ w_a(1-a))} \right) \ .
\eeq
Holding $H_0 =H(a=1)$ and $\Omega_m$ fixed, the only way to increase $D(a)$, relative to $\Lambda$CDM, near $a=1$ is for $H(a<1) < H_{\Lambda{\rm CDM}}(a<1)$. We can accomplish this by changing $w_0$, however, having smaller $H(a)$ in the past requires $w_0 < -1$, which violates the null energy condition 
(NEC)~\cite{Kontou:2020bta}.

One easy way to understand the relationship between the growth function and the NEC is through the equation
\beq
2\Mpl^2 \dot H = - (\rho + p) = -(T_{\mu \nu}X^{\mu}X^{\nu}) \ ,
\eeq
for a unit null-vector $X^{\mu}$, and stress-tensor $T_{\mu \nu}$ for a homogeneous and isotropic fluid. If we obey the NEC, $T_{\mu \nu}X^{\mu}X^{\nu} \geq 0$, the we must have $\dot H \leq 0$ and therefore, we cannot make $H(a < 1) \leq H(a=1)$.

\begin{figure}[t!]
	\centering
 \includegraphics[width=0.95\textwidth]{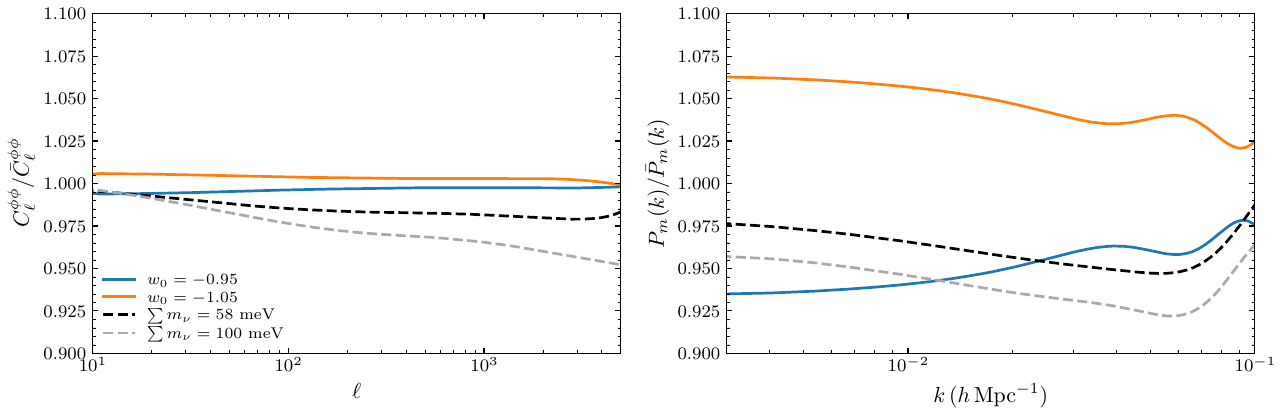}
	\caption{Fractional change to $C^{\phi\phi}_\ell$ (left) and $P_k(m)$ at $z=0$ (right) relative to $w_0=-1$ and $\sum m_\nu =0$ for different values of $w_0$ and $\sum m_\nu$. For neutrino mass, the fractional change is approximately the same size for both observables. In contrast, a 5\% change in $w_0$ has a 0.5\% level impact on CMB lensing while altering the matter power spectrum by 7\%. }
	\label{fig:DE_comb}
\end{figure}

A second challenge for a dark energy only explanation for the apparent excess clustering is that CMB lensing is most affected by $D(z)$ in the redshift range $10>z> 1$~\cite{Lewis:2006fu}. As a result, changes to the expansion history during the era of dark energy domination, $z \lesssim 0.3$, have a significantly smaller impact on the CMB lensing power spectrum than they do on $P_m(k,z=0)$, as shown in Figure~\ref{fig:DE_comb}. Specifically, we notice that neutrino mass gives rise to a few percent suppression to both the CMB lensing power spectrum and $P_m(k)$. In contrast, we see that changes to $w_0$ have an order of magnitude large impact on the matter power spectrum compared to $C_\ell^{\phi\phi}$. To put this in context, the DESI constraints on the dark energy equation of state $w_0\, (w) = -1.122^{+0.062}_{-0.054}$, with $w_a=0$ held fixed~\cite{DESI:2024mwx} (see also~\cite{DESI:2024kob}), would allow for a $10\%$ enhancement of clustering as defined by $P_m(k,z=0)$ but would have at most 1\% impact on $C_\ell^{\phi  \phi}$.


\subsection{Dynamical Dark Energy and Current Data}

Introducing dynamical dark energy is generally expected to weaken constraints on other cosmological parameters. Concretely, the Cramer-Rao bound ensures that marginalizing over a Fisher matrix with additional parameters will increase the variance of any parameter, relative to the measurement with additional parameters fixed. However, when we include priors that restrict the parameter space, like $\sum m_\nu > 0$ or the NEC ($w\geq -1$), the 95\% confidence intervals can appear to be more strongly constrained due to shifts in the central value of the parameter, rather than due to a decrease in the variance. In this sense, the pattern of unusually strong constraints on $\sum m_\nu$ and/or a more stringent constraint when marginalizing over additional parameters, suggests that the data favors a shift towards negative values. Allowing $\sum m_\nu < 0$ allows us to explore whether this is indeed relevant to analyses of models with both $\sum m_\nu$ and dynamical dark energy. 

\begin{figure}[t!]
    \centering
    \includegraphics[width=0.6\textwidth]{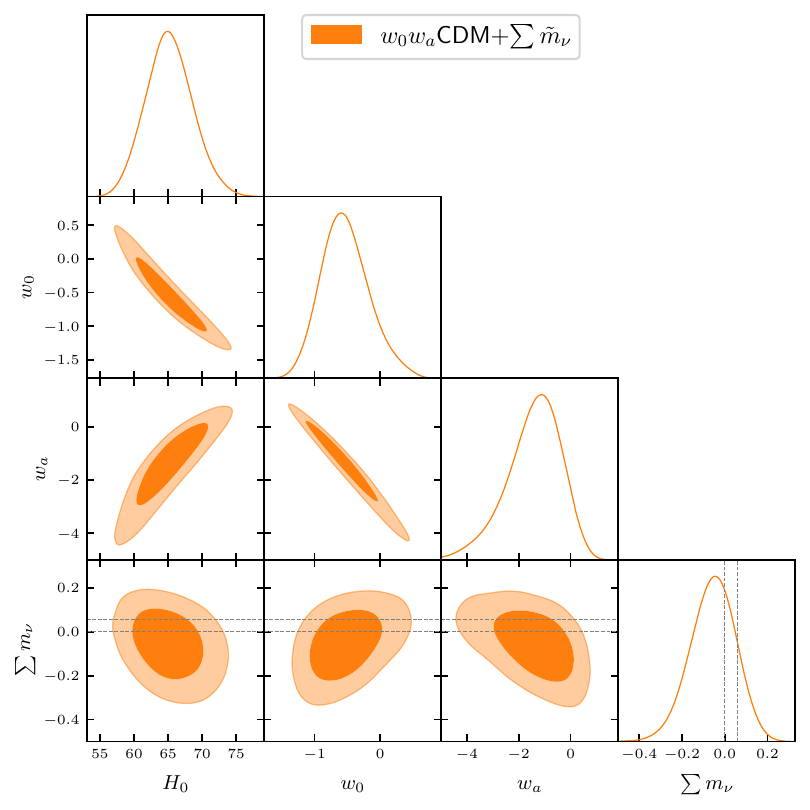}
    \caption{Triangle plot showing parameter constraints from Planck + ACT lensing + DESI BAO in the $w_0w_a$CDM+$\sum \tilde{m}_\nu$ model.   Dashed lines show vanishing neutrino mass $\sum m_\nu = 0$ and the minimal sum of neutrino mass $\sum m_\nu = 58$~meV. Values of neutrino mass are reported in eV and $H_0$ in $\mathrm{km}\, \mathrm{s}^{-1}\, \mathrm{Mpc}^{-1}$.}
    \label{fig:triangle_DE}
\end{figure}

The results of analyzing CMB+DESI data for $\Lambda\mathrm{CDM}+\sum m_\nu+w_0+w_a$ are shown in Figure~\ref{fig:triangle_DE}. As expected, there is a significant degeneracy between neutrino mass and dark energy that significantly weakens the constraints on $\sum m_\nu$. However, the freedom to explain the signal with dark energy does not remove the preference for negative neutrino mass; the maximum likelihood remains negative although shifted slightly towards positive values. Moreover, the regions of parameter space where the neutrino masses are more positive correspond to $w_0 > -0.5$ and $w_a  < -2$ which implies that very large departures from $\Lambda$CDM are needed to move the behavior from the neutrino sector to dark energy. In addition, these regions prefer smaller values of $H_0$ putting the models in further tension with the local measurements.

It is additionally noteworthy that the best-fit values of $w_0$ and $w_a$ also are largely unchanged when introducing neutrino mass while allowing $\sum m_\nu < 0$. The contours in Figure~\ref{fig:triangle_DE} correspond to measurement of $w_0 = -0.53^{+0.32}_{-0.41}$ and $w_a = -1.42^{+1.2}_{-0.81}$. Compare these to the DESI results for $\Lambda\mathrm{CDM}+w_0+w_a$ of $w_0 = -0.45^{+0.34}_{-0.21}$  and $w_a = -1.79^{+0.48}_{-1.0}$~\cite{DESI:2024mwx}. While the results with $\sum m_\nu$ fixed have smaller uncertainties, the maximum likelihood occurs at roughly the same values of all three parameters.

These results hardly suggest a dark energy interpretation of the enhanced clustering signal. If the preference for $w_a \neq 0$ and $w_0 \neq -1$ from DESI BAO was consistent with $\sum m_\nu \geq 58$~meV, we would generally expect a significant upward shift in the maximum likelihood values when including dynamical dark energy. This behavior was anticipated at the end of the previous section: because the CMB lensing kernel receives significant contributions for matter over-densities in the redshift range $1<z< 10$, changes to the expansion history at $z \lesssim 2$ have a much larger impact on the BAO than they will on lensing.

\subsection{Non-Phantom Dynamical Dark Energy}

The results with $w_0$ and $w_a$ also point to the challenges in explaining $\sum m_\nu < 0$ within the space of concrete dark energy models. For example, there is good reason to expect the null energy condition to hold~\cite{Rubakov:2014jja,Joyce:2014kja,Kontou:2020bta}, in which case we would focus on the so called Non-Phantom Dynamical Dark Energy (NPDDE) regime of $w_0$ and $w_a$. This is a class of models where the dark energy is allowed to evolve in time, but subject to the condition that the equation of state of dark energy $w(z)\geq -1$. As we saw in the previous section, $w(z) \geq -1$ will change the growth function $D(z)$ but in a way that will suppress growth (at fixed $H_0$ and $\Omega_m$) and hence the amplitude of the matter power spectrum. This is still just a $w_0$ and $w_a$ parameterization, but restricted to points that can be achieved in many physical examples. 

\begin{figure}[ht!]
    \centering
    \includegraphics[width=0.6\textwidth]{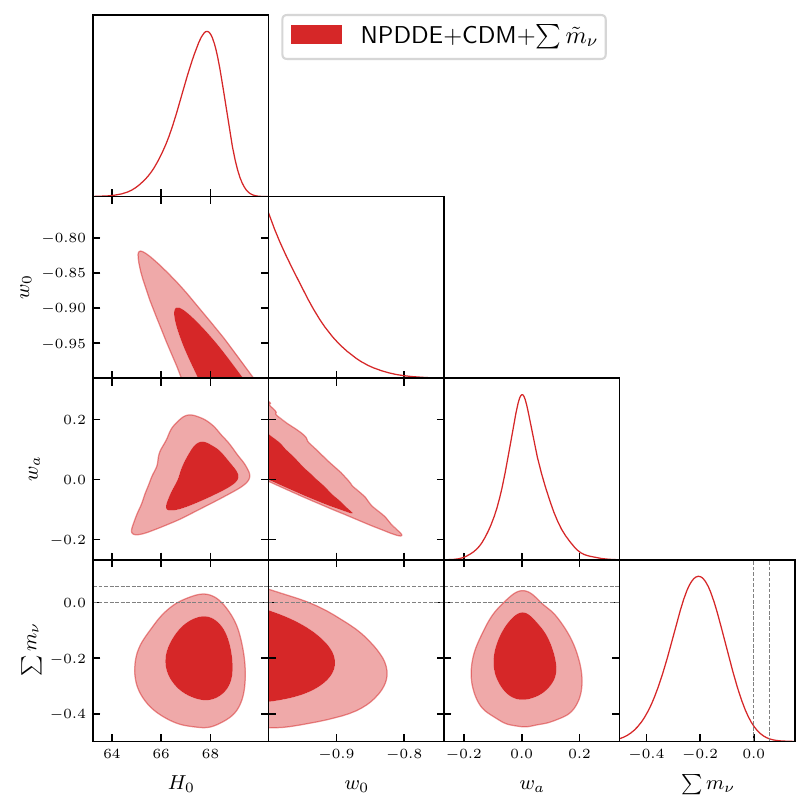}
    \caption{Triangle plot showing parameter constraints from Planck + ACT lensing + DESI BAO in the NPDDE+CDM+$\sum \tilde{m}_\nu$ model.   Dashed lines show vanishing neutrino mass $\sum m_\nu = 0$ and the minimal sum of neutrino mass $\sum m_\nu = 58$~meV. Values of neutrino mass are reported in eV and $H_0$ in $\mathrm{km}\, \mathrm{s}^{-1}\, \mathrm{Mpc}^{-1}$.}
    \label{fig:triangle_NPDDE}
\end{figure}

However, as we saw earlier, the parameter space of NPDDE corresponds precisely to the regime where we would expect a decrease in clustering, rather than the increase that is preferred by the data. As a result, we might expect that marginalizing over NPDDE will push $\sum m_\nu$ towards even more negative values, in order to maintain the amplitude of the CMB lensing power spectrum. The results for $\sum  m_\nu$ with and without NPDDE are shown in Figure~\ref{fig:triangle_NPDDE} and Table~\ref{tab:constraints}. As anticipated, we see that $\sum m_\nu$ is pushed to lower values. Specifically from general $w_0 w_a$ to NPDDE, we have $\sum m_\nu = - 60^{+110}_{-100} \to -210\pm 97$ meV. This confirms the expectation that we need to violate the null energy condition in order to increase the clustering of matter and thus reduce the preference for negative neutrino mass.

This behavior also explains why previous analyses for NPDDE with physical $\sum m_\nu$ found {\it stronger} bounds than in $\Lambda$CDM~\cite{Vagnozzi:2018jhn,Berghaus:2024kra}. We know that marginalizing over $w_a$ and $w_0$ must increase $\sigma^2(\sum m_\nu)$. However, imposing the non-phantom condition gives rise to such a large shift towards negative values that the upper limit on the neutrino mass is stronger than if $w_a$ and $w_0$ were held fixed. This pattern is also present for physical $\sum m_\nu \geq 0$ and gives rise to more stringent 95\% upper limits, even if we don't directly include the negative mass regime. This is just one of several examples where previous analyses hinted at the preference for negative neutrino mass even without directly extending the models to negative values.

\begin{table}[t!]
    \begin{center}
    
\resizebox{\textwidth}{!}{
\begin{tabular} { l  d f e g}
\noalign{\vskip 3pt}\hline\noalign{\vskip 1.5pt}\hline\noalign{\vskip 5pt}
 \multicolumn{1}{c}{\bf } &  \multicolumn{1}{c}{\bf $\Lambda$CDM+$\sum m_\nu$} &  \multicolumn{1}{c}{\bf $\Lambda$CDM+$\sum \tilde{m}_\nu$} &  \multicolumn{1}{c}{\bf $w_0w_a$CDM+$\sum \tilde{m}_\nu$} &  \multicolumn{1}{c}{\bf NPDDE+CDM+$\sum \tilde{m}_\nu$}\\
\noalign{\vskip 3pt}\cline{2-5}\noalign{\vskip 3pt}

 Parameter &  68\% limits &  68\% limits &  68\% limits &  68\% limits\\
\hline
{\boldmath$\log(10^{10} A_\mathrm{s})$} & $3.051\pm 0.014            $ & $3.030\pm 0.017            $ & $3.032^{+0.017}_{-0.015}   $ & $3.029\pm 0.017            $\\

{\boldmath$n_\mathrm{s}   $} & $0.9692\pm 0.0037          $ & $0.9708\pm 0.0038          $ & $0.9669\pm 0.0046          $ & $0.9729\pm 0.0040          $\\

{\boldmath$100\theta_\mathrm{MC}$} & $1.04112\pm 0.00029        $ & $1.04118\pm 0.00029        $ & $1.04100\pm 0.00032        $ & $1.04128\pm 0.00030        $\\

{\boldmath$\Omega_\mathrm{b} h^2$} & $0.02249\pm 0.00013        $ & $0.02255\pm 0.00014        $ & $0.02243\pm 0.00016        $ & $0.02262\pm 0.00014        $\\

{\boldmath$\Omega_\mathrm{c} h^2$} & $0.11852\pm 0.00088        $ & $0.11780\pm 0.00097        $ & $0.1194\pm 0.0014          $ & $0.1170\pm 0.0011          $\\

{\boldmath$\tau_\mathrm{reio}$} & $0.0585\pm 0.0074          $ & $0.0510\pm 0.0083          $ & $0.0499^{+0.0083}_{-0.0074}$ & $0.0515\pm 0.0082          $\\

{\boldmath$ \sum m_\nu            $} & $< 0.0326                  $ &                              &                              &                             \\

{\boldmath$\sum \tilde{m}_\nu    $} &  & $-0.156^{+0.093}_{-0.085}  $ & $-0.06^{+0.11}_{-0.10}     $ & $-0.210\pm 0.097           $\\

{\boldmath$w_0              $} & & & $-0.53^{+0.32}_{-0.41}     $ & $< -0.936                  $\\

{\boldmath$w_a             $} & & & $-1.42^{+1.2}_{-0.81}      $ & $0.012^{+0.065}_{-0.075}   $\\

\hline

$A_\mathrm{s}              $ & $\left(\,2.114^{+0.026}_{-0.029}\,\right)\cdot 10^{-9}$ & $\left(\,2.070\pm 0.035\,\right)\cdot 10^{-9}$ & $\left(\,2.073\pm 0.035\,\right)\cdot 10^{-9}$ & $\left(\,2.068\pm 0.035\,\right)\cdot 10^{-9}$\\

$H_0                       $ & $68.33\pm 0.43             $ & $68.87\pm 0.45             $ & $65.2\pm 3.5               $ & $67.5^{+1.1}_{-0.69}       $\\

$\Omega_\mathrm{m} h^2     $ & $0.14131\pm 0.00084        $ & $0.14036\pm 0.00092        $ & $0.1418\pm 0.0013          $ & $0.1396\pm 0.0010          $\\

$\sigma_8                  $ & $0.8175^{+0.0076}_{-0.0058}$ & $0.8123\pm 0.0078          $ & $0.796\pm 0.029            $ & $0.793^{+0.015}_{-0.011}   $\\
\hline
\end{tabular}
}
\end{center}
\caption{Parameter constraints from Planck + ACT lensing + DESI BAO in $\Lambda\mathrm{CDM}$ plus physical neutrino mass (blue), $\Lambda\mathrm{CDM}$ plus parameterized neutrino mass (green), $w_0w_a\mathrm{CDM}$ plus parametrized neutrino mass (orange), and $\mathrm{NPDDE}+\mathrm{CDM}$ plus parameterized neutrino mass (red).  All constraints are given as $68\%$ limits.  Values of neutrino mass are reported in eV and $H_0$ in $\mathrm{km}\, \mathrm{s}^{-1}\, \mathrm{Mpc}^{-1}$.  
}
\label{tab:constraints}
\end{table}

\section{The Future of Neutrino Mass Measurements}\label{sec:future}

The state of the current measurement of $\sum m_\nu$ suggests the preference for $\sum m_\nu < 0$ is likely to persist for the foreseeable future. We saw in Section~\ref{sec:lensing} that shifts in the optical depth or BAO data due to different likelihoods and/or datasets do not shift the parameters sufficiently to meaningfully impact the conclusion. In Section~\ref{sec:darkenergy}, we saw that this preference was also robust to changes in the late-time expansion history as well.

Naturally, we would like to know what observations might clarify our understanding of cosmological neutrino mass and clustering in the universe. DESI is still collecting data, as are numerous ground-based CMB surveys~\cite{Chang:2022tzj}. In this section, we will address whether it is possible this future data could be consistent with $\sum m_\nu \geq 58$ meV without being inconsistent with the current Planck and DESI data. In addition, other discrepancies between different types of cosmological data are already known, and we assess the possibility that resolving these tensions could shed light on neutrino mass, or vice versa.

\subsection{Future BAO Data}

The DESI year one BAO data was extracted the spectroscopic redshifts of 6.4 million extragalactic sources~\cite{DESI:2024mwx}. This is only a quarter of the expected 23.6 million objects that will form the five-year survey~\cite{DESI:2016fyo}. One might imagine that the inclusion of this data could have sufficient statistical power to drive the measurement of $\sum m_\nu$ back to the expected range. However, as emphasized above, the measurement of $\Omega_m h^2$ with DESI is not particularly inconsistent with the CMB. As a result, one expects that future DESI data would have to generate a very large shift, not only away from the current DESI central value, but even the CMB-only measurements of $\Omega_m$. In this sense, DESI data that is consistent with the CMB is unlikely to display evidence for a positive sum of neutrino masses.

In order to test this idea, we generate mock DESI 5-year BAO data to combine with existing CMB data. To be consistent with the Planck uncertainties on cosmological parameters, we randomly generated a fiducial cosmology from the Planck TTTEEE $\Lambda$CDM (with $\sum m_\nu = 58$ meV held fixed) posterior distribution. From this fiducial cosmology, we then generated simulated DESI 5-year BAO data based on the full survey volume and number density. By construction, the BAO data should be consistent with the Planck best-fit cosmology with $\sum m_\nu > 0$ without including the lensing power spectrum.

\begin{figure}[t!]
    \centering
    \includegraphics[width=0.8\textwidth]{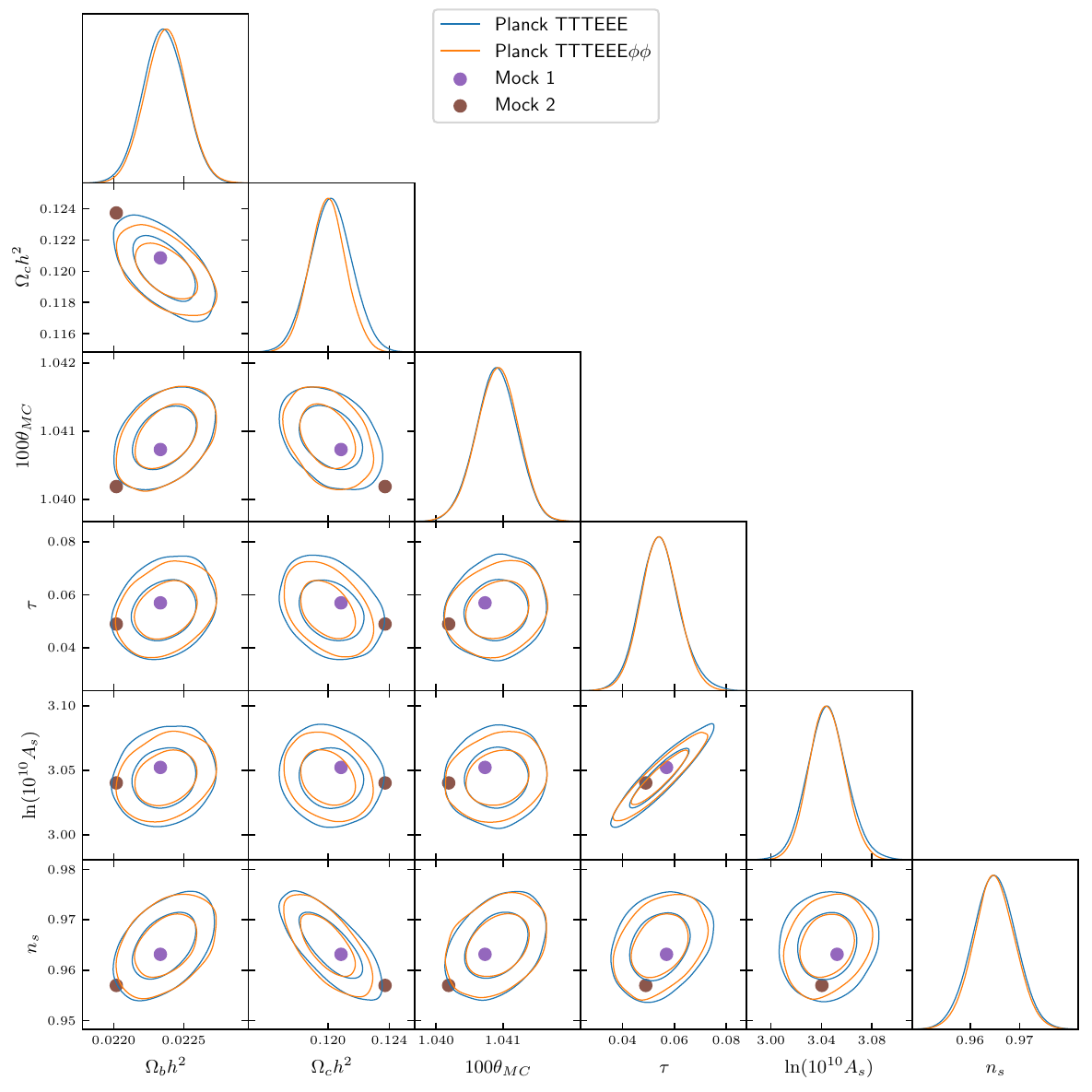}
    \caption{Triangle plot showing cosmological parameters used to generate the mock DESI data overlaid on the Planck 2018 $\Lambda$CDM constraints~\cite{Planck:2018vyg}.   }
    \label{fig:triangle_Planck}
\end{figure}

The purpose of the mock data is to test the possibility that further DESI data will resolve the current neutrino mass tension in the data. If the tension is due entirely to the current BAO data being inconsistent with the CMB, then generating the mock data from a fiducial $\Lambda$CDM cosmology  consistent with the CMB should produce measurements consistent with $\sum m_\nu \geq 58$ meV. On the other hand, if we continue to find $\sum m_\nu < 0$ when we include the mock BAO data and lensing, it would then imply that the lensing data is inconsistent the Planck TTTEEE best-fit cosmology. In this regard, by taking the BAO data from the CMB cosmology, we are testing the consistency of the lensing potential with TTTEEE at the sensitivity of Planck + DESI. Second, by including mock data for the full 5-year survey, we can see how much effect better BAO data can have on $\sum m_\nu$ even if there is some discrepancy between the lensing potential amplitude and the fiducial cosmology. Specifically, if the BAO data was providing important information about $\sum m_\nu$ directly, then the full BAO data could move $\sum m_\nu$ to positive values even if the lensing potential was in conflict.

\begin{figure}[t!]
    \centering
    \includegraphics[width=0.6\textwidth] {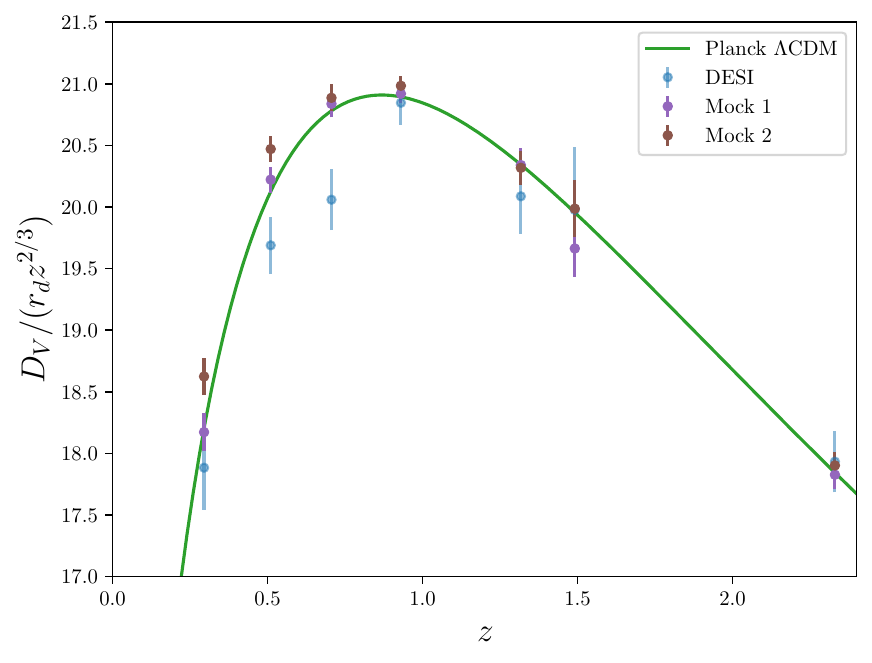}
    \caption{Mock 5-year DESI measurements of $D_v/r_d$ overlaid on the real 1-year DESI measurements.  The cosmological parameters used to generate the mock data are shown in Figure~\ref{fig:triangle_Planck}.}
    \label{fig:DV_over_rs_mocks}
\end{figure}

We generated two Mocks (1 \& 2) with fiducial $\Lambda$CDM cosmological parameters drawn from the Planck TTTEEE posterior distribution with a fixed $\sum m_\nu = 58$ meV. The values of these parameters, shown in Figure~\ref{fig:triangle_Planck}, where chosen to represent roughly 1$\sigma$ and 2$\sigma$ upward fluctuations of $\Omega_m$ relative to the maximum posterior cosmology in order favor large (more positive) values of $\sum m_\nu$ in the mocks (given that the current values of $\Omega_m$ from DESI BAO gives rise to negative neutrino mass. Given these cosmologies, we generated DESI 5-year BAO data consistent with the fiducial $\Lambda$CDM cosmology and the forecasted uncertainties. It is important to note that the mock DESI data is being generated in such a way that it is consistent with Planck TTTEEE and $\Lambda$CDM. Moreover, the values of the cosmological parameters have been post-selected to be more likely to give positive neutrino masses. The mock data is shown in Figure~\ref{fig:DV_over_rs_mocks} and Table~\ref{tab:DESI_mocks} and can be seen to be a large upward fluctuation of the BAO data in a number of redshift bins.

\begin{table}[t!]
\centering
\begin{tabular}{lrrrrrrr}
\noalign{\vskip 3pt}\hline\noalign{\vskip 1.5pt}\hline\noalign{\vskip 5pt}
Redshift $(z)$                                      & 0.295 & 0.510 & 0.706 & 0.930 & 1.317 & 1.491 & 2.33  \\
\noalign{\vskip 3pt}\cline{2-8}\noalign{\vskip 3pt}
\rowcolor{colorC0!10} 
DESI $D_v/r_d$                           & 7.93  & 12.6  & 15.9  & 19.9  & 24.1  & 26.1  & 31.5  \\
\rowcolor{colorC0!10} 
DESI $\sigma(D_v/r_d)$                   & 0.15  & 0.15  & 0.20  & 0.17  & 0.36  & 0.67  & 0.44  \\
\noalign{\vskip 3pt}
\rowcolor{colorC4!10} 
Mock 1 $D_v/r_d$                         & 8.05  & 12.9  & 16.5  & 19.9  & 24.4  & 25.7  & 31.3  \\
\rowcolor{colorC4!10} 
Mock 1 $\Delta(D_v/r_d)/\sigma(D_v/r_d)$ & 0.85  & 2.3   & 3.1   & 0.42  & 0.84  & -0.62 & -0.42 \\
\noalign{\vskip 3pt}
\rowcolor{colorC5!10} 
Mock 2 $D_v/r_d$                         & 8.25  & 13.1  & 16.6  & 20.0  & 24.4  & 26.1  & 31.5  \\
\rowcolor{colorC5!10} 
Mock 2 $\Delta(D_v/r_d)/\sigma(D_v/r_d)$ & 2.17  & 3.36  & 3.34  & 0.77  & 0.76  & 0.019 & -0.12
\\
\hline
\end{tabular}
\caption{Table showing the DESI measurements of $D_v/r_d$ and associated uncertainties.  The mock BAO samples and the shifts relative to current DESI data (in units of current DESI uncertainties) are also shown.}
\label{tab:DESI_mocks}
\end{table}

\begin{figure}[ht!]
    \centering
    \includegraphics[width=0.8\textwidth]{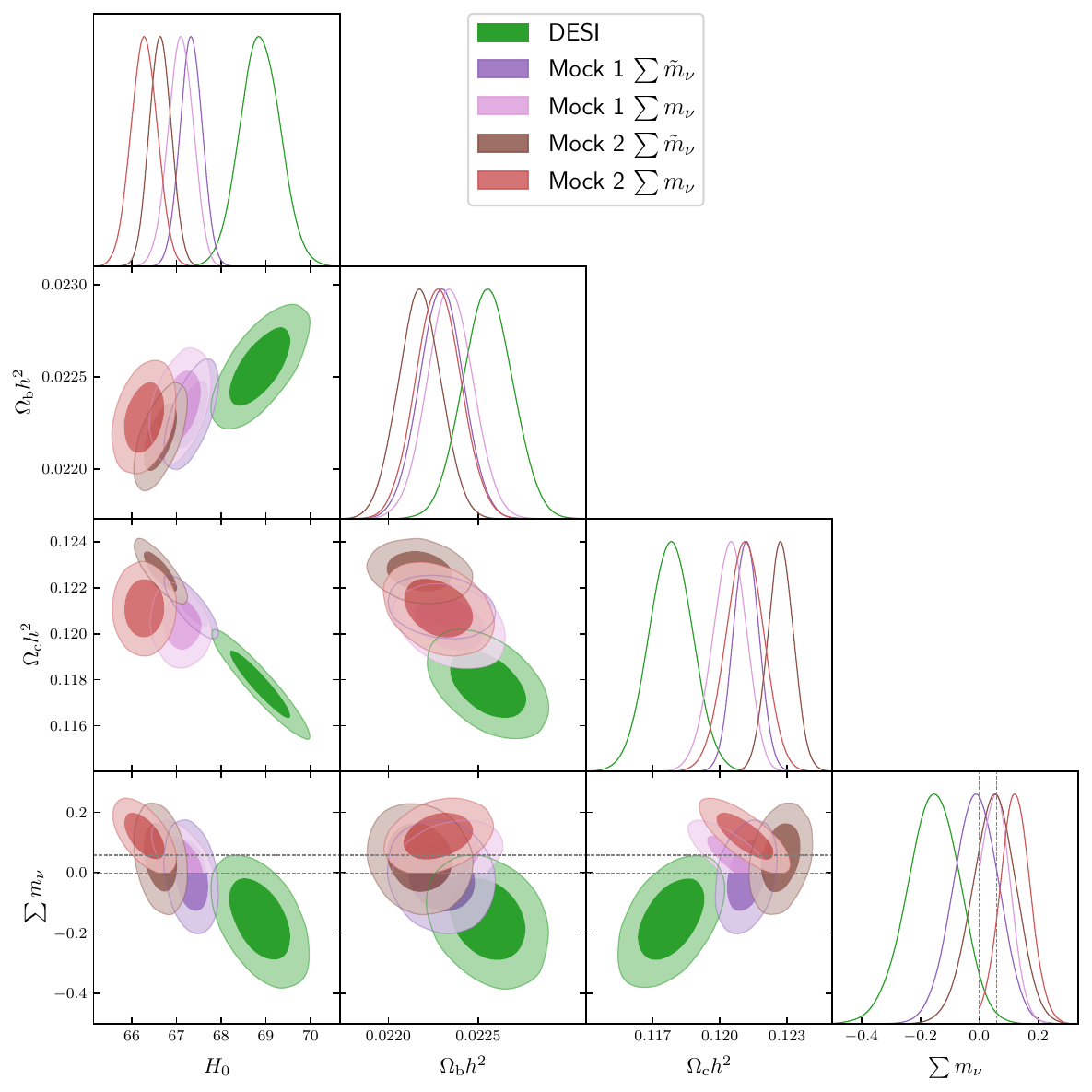}
    \caption{Triangle plot showing parameter constraints from Planck + ACT lensing + Mock DESI.  The cosmological parameters used to generate the mock DESI data were drawn from a distribution consistent with Planck 2018 constraints as shown in Figures~\ref{fig:triangle_Planck} and \ref{fig:DV_over_rs_mocks}.   Dashed lines show vanishing neutrino mass $\sum m_\nu = 0$ and the minimal sum of neutrino mass $\sum m_\nu = 58$~meV. Values of neutrino mass are reported in eV and $H_0$ in $\mathrm{km}\, \mathrm{s}^{-1}\, \mathrm{Mpc}^{-1}$.}
    \label{fig:triangle_mocks}
\end{figure}

The analysis of the Placnk TTTEEE+$\phi \phi$ with the mock data is shown in Figure~\ref{fig:triangle_mocks}. By construction, we see that the values of $\Omega_c h^2$ in the posterior is substantially larger than with Plank + DESI. Mock 1, which represents a 0.47$\sigma$ upward fluctuation in $\Omega_c h^2$ relative to the Planck TTTEEE best-fit $\Lambda$CDM cosmology and a 2.8$\sigma$ upward fluctuation of $\Omega_m$ compared to the best-fit CMB+DESI $\Lambda$CDM cosmology, remains peaked at $\sum \tilde{m}_\nu < 0$ but does favor a non-zero and positive physical neutrino mass. Mock 2, which represents a 2.5$\sigma$ upward fluctuation in $\Omega_c h^2$ relative to Planck and a 5.9$\sigma$ upward fluctuation of $\Omega_m$ relative to CMB+DESI, is consistent with positive neutrino mass in both $\sum \tilde m_\nu$ and $\sum m_\nu$ and thus would represent a possible detection of neutrino mass with the Planck + DESI 5-year data.

It is noticeable that the posterior distributions of $\sum \tilde m_\nu$ and $\sum m_\nu$ differ significantly when analyses with both mocks. In particular, the $\sum \tilde m_\nu$ analyses are shifted towards smaller values of neutrino mass than $\sum m_\nu$. It is important that $\sum m_\nu$ contains the additional physical effect of neutrino mass on the expansion rate; the mock data is constructed to be consistent with $\sum m_\nu = 58$ meV and therefore should prefer positive mass. In contrast, $\sum \tilde m_\nu$ only includes the effect of neutrions on CMB lensing, and therefore the BAO data has no direct impact on $\sum \tilde m_\nu$. We can interpret the shift between the $\sum m_\nu$ and $\sum \tilde m_\nu$ analyses as an additional sign that the CMB lensing amplitude is larger than would be expected given the cosmological parameters determined by a $\Lambda$CDM cosmology consistent with the primary CMB.

Using only the Planck data, it entirely possible that the CMB lensing amplitude is consistent with $\Lambda$CDM. However, from this analysis, we see that the values of the fiducial cosmological parameters that would be needed to allow $\sum m_\nu \geq 58$ meV are consistent at 2$\sigma$ with Planck TTTEEE (i.e.~without including the lensing data) but in strong (approximately $6\sigma$) tension with DESI year 1 and eBOSS BAO data. A detection of $\sum m_\nu>0$ with DESI 5-year data would therefore require the current DESI distance data to have been systematically low by several $\sigma$ in multiple redshift bins (see Figure~\ref{fig:DV_over_rs_mocks} and Table~\ref{tab:DESI_mocks}). Barring a major systematic error the DESI year 1 BAO analysis, it is unlikely that additional DESI data will shift $\sum m_\nu$ to positive values.

\subsection{Future CMB Lensing}\label{sec:flens}

The signal of a negative neutrino mass, as identified in this paper in terms of CMB+BAO, is a larger than expected amplitude of CMB lensing. As such, we might wonder how future measurements of the CMB, particularly with regards to lensing, will impact our understanding of the neutrino mass. From a number of previous studies, it is known that the improving the CMB lensing maps beyond Planck has little influence on $\sigma(\sum m_\nu)$. With DESI 5-year BAO, the measurement of neutrino mass is expected to be limited entirely by the degeneracy with the optical depth $\tau$~\cite{Font-Ribera:2013rwa,Allison:2015qca,CMB-S4:2016ple,Green:2021xzn}. 

The value of additional CMB lensing data is therefore driven by the added value of consistency between surveys and robustness to systematics. Current measurements of lensing are driven by TT lensing reconstruction with Planck~\cite{Planck:2018lbu,Carron:2022eyg} and ACT~\cite{ACT:2023dou,ACT:2023kun}.  The Planck reconstruction measures the amplitude of lensing with 2.6\% precision using the minimum variance combination of temperature and polarization while temperature-only reconstruction gives 3.3\% precision and polarization-only measurement provides 11\% precision~\cite{Planck:2018lbu,Carron:2022eyg}. ACT measures the lensing amplitude at 2.3\% precision in their minimum variance analysis, at 3.7\% using only temperature, and at 4.9\% using only polarization. The analysis with and without the addition of ACT lensing data makes little difference for constraints on neutrino mass, as can be seen in Figure~\ref{fig:mnu_1d_lensing_variants}.  The consistency between the Planck and ACT lensing analyses is non-trivial, since the surveys are complementary. The Planck lensing measurement utilizes 60\% of the sky while ACT covers about 23\% of the sky with lower noise and a smaller beam, meaning that the two surveys are sensitive to overlapping, but not identical, lensing modes. 

An analysis of only the Planck CMB power spectra, without lensing reconstruction, provides a measurement of the effects of CMB lensing (through peak smoothing) at a precision of about 6\%~\cite{Planck:2018lbu}.  This two-point CMB lensing analysis favors a lensing power that is larger than expectations from $\Lambda$CDM at almost 3$\sigma$.  However, when combined with lensing reconstruction, the lensing amplitude is consistent with expectations at 2$\sigma$.  As discussed in Section~\ref{subsec:2pt4pt}, the extra peak smoothing is not primarily responsible for driving the preference for negative neutrino mass.

\begin{figure}[t!]
    \centering
    \includegraphics[width=0.7\textwidth]{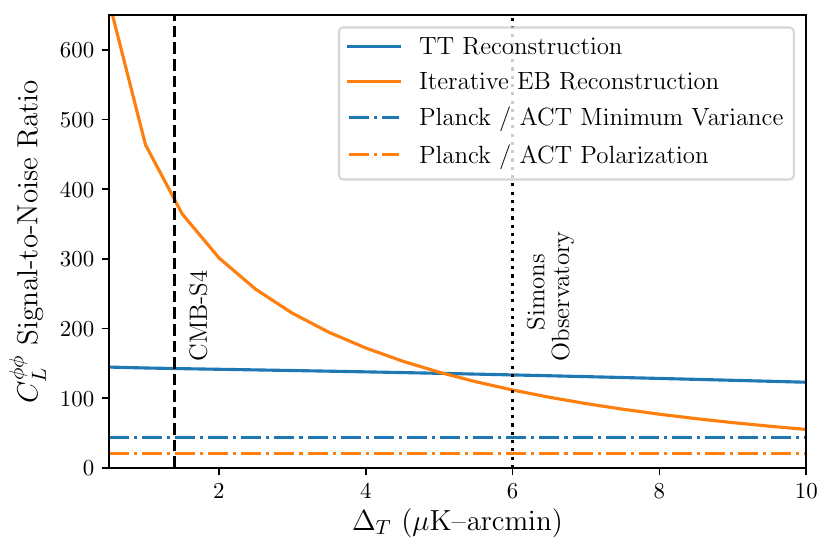}
    \caption{Signal-to-noise ratio of CMB lensing reconstruction as a function of map depth assuming CMB experiments with a beam size of $1.4$~arcmin and a survey with $f_\mathrm{sky} = 0.5$.  These estimates assume $30<\ell<3000$ for temperature (excluding the small-scale modes expected to be most contaminated by extragalactic foregrounds) and $30<\ell<5000$ for polarization. We include the improvement from iterative delensing as implemented in Ref.~\cite{Hotinli:2021umk}. The current signal-to-noise ratio of the lensing reconstruction measurement using data from Planck~\cite{Planck:2018lbu,Carron:2022eyg} and ACT~\cite{ACT:2023dou,ACT:2023kun} are shown as horizontal lines for each of the minimum variance estimate and the polarization-only reconstruction.  We show only a single line for Planck and ACT since the lensing signal-to-noise from the two surveys is similar (the measurement precision is listed for each individually in the main text).  Vertical lines indicate the approximate noise levels expected from Simons Observatory~\cite{SimonsObservatory:2018koc} and CMB-S4~\cite{CMB-S4:2016ple}.}
    \label{fig:lensing_SN}
\end{figure}

Future surveys have the potential to change the nature of the neutrino mass in two important ways. First, with lower noise levels, the measurement of the lensing potential from E- to B-mode conversion will surpass the information available from TT reconstruction~\cite{SimonsObservatory:2018koc,CMB-S4:2016ple}, as can be seen in Figure~\ref{fig:lensing_SN}. The TT power spectra are known to be limited by unresolved point sources (foregrounds). This sets a noise floor for TT lensing reconstruction at high $\ell$, and furthermore one might worry that non-Gaussianity in these point sources could contaminate the lensing reconstruction~\cite{vanEngelen:2013rla,Ferraro:2017fac}. Because point sources are known to be polarized at the 1-percent level, lensing reconstruction using the EB lensing estimator is expected to be more robust against the effects of foreground contamination.

The second benefit of reconstructing maps of the CMB lensing field at high signal to noise is that it enables the delensing of the CMB maps~\cite{Seljak:2003pn,Green:2016cjr,Hotinli:2021umk}.  Delensing has numerous advantages, including allowing for tighter constraints on cosmological parameters~\cite{Green:2016cjr,Hotinli:2021umk,Ange:2023ygk,Trendafilova:2023xtq}.  Furthermore, delensing allows the possibility of disentangling the physical effects that impact the primary CMB anisotropies and the matter fluctuations that cause CMB lensing.  This cleaner separation of the effects of CMB lensing would allow for a more careful assessment of what aspects of the CMB data contribute to the preference for negative neutrino mass. 

While the cosmological constraining power is currently weaker, optical weak lensing surveys are approaching statistical uncertainty comparable to CMB lensing~\cite{Heymans:2020gsg,DES:2021wwk,Li:2023tui,Dalal:2023olq}. Weak lensing studies with upcoming galaxy surveys~\cite{LSSTScience:2009jmu,Euclid:2024yrr} may soon provide an additional useful source of information to probe matter clustering and the cosmological measurement of neutrino mass.


\subsection{Relation to $S_8$ and Hubble Tensions}

The cosmological measurement of neutrino mass necessarily requires combining data from different sources. However, combining data in this way can be particularly sensitive to calibration issues and/or new physics that might not be apparent in the data from individual surveys on their own. Tensions between cosmic surveys have been a lingering issue~\cite{Abdalla:2022yfr}, including the discrepancy of the local measurement of $H_0$~\cite{Riess:2021jrx} and the inferred value from CMB+BAO, and the measurement of $S_8$ (and/or $\sigma_8$) from large scale structure~\cite{DES:2021wwk,Busch:2022pcx,Chen:2024vuf} and the inferred value from the CMB. While in principle, these issues could be unrelated, one should take seriously that they could all follow from a single underlying cause. Of course, DESI itself will have something to contribute to our understanding of these tensions, particularly $S_8$, via full shape and cross-correlations analyses, such as~\cite{Sailer:2024coh,Kim:2024slb,Chen:2024vvk}.

\begin{figure}[t!]
    \centering
    \includegraphics[width=0.58\textwidth]{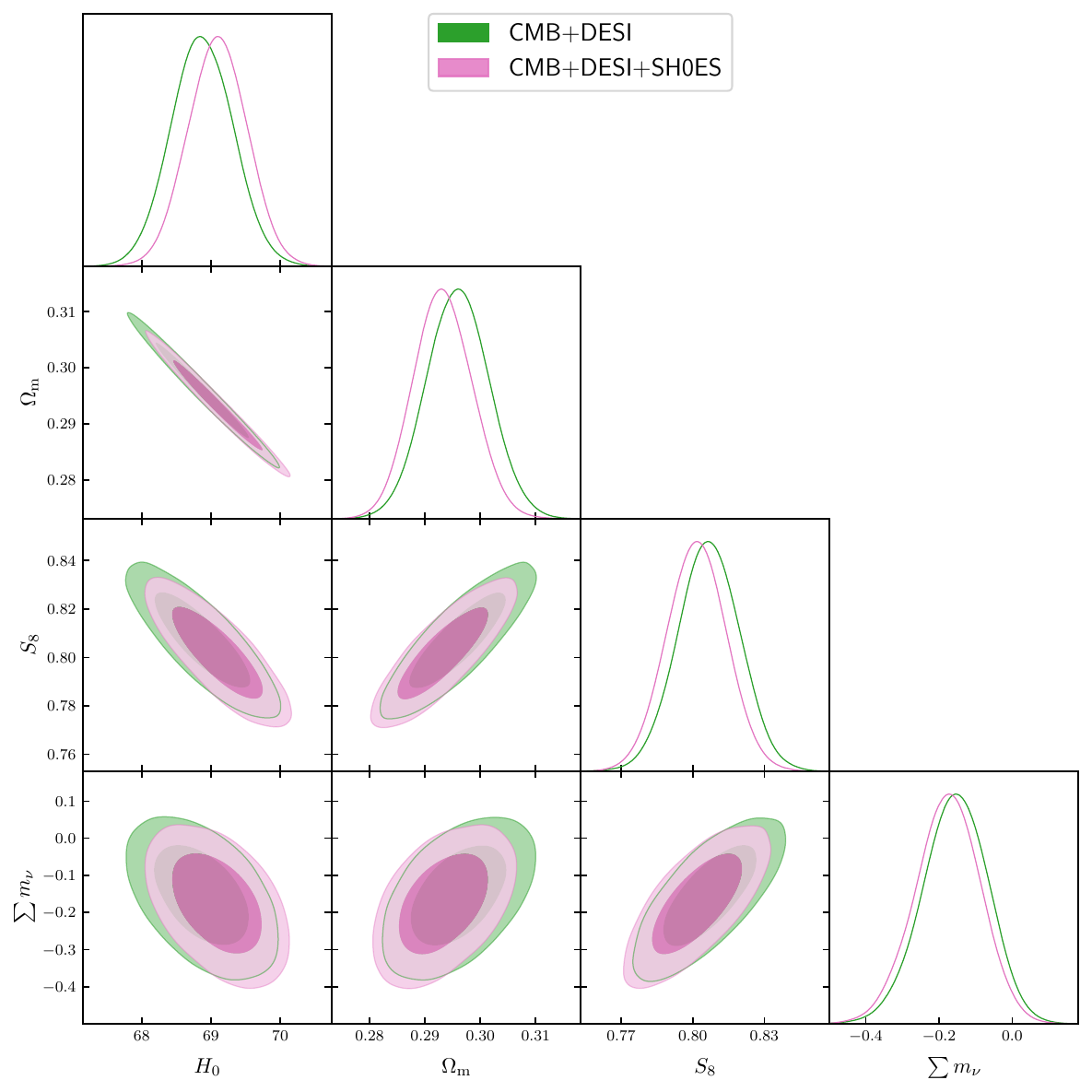}
    \caption{Triangle plot showing parameter constraints from Planck + ACT lensing + DESI BAO + SH0ES in the $\Lambda$CDM+$\sum \tilde{m}_\nu$ model.    Dashed lines show vanishing neutrino mass $\sum m_\nu = 0$ and the minimal sum of neutrino mass $\sum m_\nu = 58$~meV. Values of neutrino mass are reported in eV and $H_0$ in $\mathrm{km}\, \mathrm{s}^{-1}\, \mathrm{Mpc}^{-1}$.}
    \label{fig:triangle_SHOES_2}
\end{figure}

Interestingly, the degeneracies between $\sum \tilde m_\nu$, $S_8$, and $H_0$ are such that negative neutrino mass also removes some of the existing tension, as shown in Figure~\ref{fig:triangle_SHOES_2}. Specifically, DESI prefers lower values of $\Omega_m$, which would be inconsistent with the larger lensing amplitude in $\Lambda$CDM. The negative neutrino mass corrects for the lensing amplitude and allows for smaller values of $\Omega_m$. Our implementation of negative neutrino mass leaves the matter power spectrum unchanged for varying $\sum \tilde{m}_\nu$, and thus we find in the regions with negative neutrino mass and lower $\Omega_m$, the values of $\sigma_8$ and $S_8$ are reduced as well. Meanwhile, lower values of $\Omega_m$ also favor higher values of $H_0$.  Constraints from the CMB power spectra exhibit an approximate degeneracy for $\Omega h^3 = \mathrm{const}$~\cite{Planck:2016tof}.  This can be understood since a decrease in the matter density leads to an increase in the size of the sound horizon, which for fixed $\theta_s$ (very well constrained through CMB spectrum peak spacing) requires an increased distance to the surface of last scattering, thereby leading to a preference for increased $H_0$~\cite{Planck:2016tof,Knox:2019rjx}.

While the allowing for negative neutrino mass reduces the tension between the CMB and the local $H_0$ measurement, the inclusion of the distance ladder measurement from SH0ES does little to alter results.  In Figure~\ref{fig:triangle_SHOES_2}, we show constraints derived from combining CMB and DESI BAO data with data from SH0Es, implemented here through constraints on the intrinsic brightness~\cite{Riess:2020fzl} of Pantheon supernovae~\cite{Pan-STARRS1:2017jku} (not a direct constraint on the value of $H_0$). As we saw in Section~\ref{sec:darkenergy}, the preference for negative neutrino mass is not driven by the expansion history. As a result, the preference a modestly larger value of $H_0$ with Planck + DESI is driven by $\Omega_m$ and not the expansion around $z\approx 0$. As a result, we do not see a meaningful shift in any parameter with the addition of supernova data. However, it is possible that beyond the Standard Model physics could offer a common resolution to both the Hubble tension and negative neutrino mass~\cite{Lynch:2024hzh}.

\section{Conclusions}\label{sec:conclusion}

The measurement of neutrino mass is one of the most anticipated results of the coming generation of galaxy and CMB surveys. Both types of data play an integral part of determining the absolute scale of neutrino mass and the combination of sensitivities is reaching the point where a 2-5$\sigma$ measurement of $\sum m_\nu \gtrsim 58$ meV was potentially achievable.

The trend in current data for $\sum m_\nu < 0$ significantly changes the landscape of what current and future surveys will be able to tell us about the masses of neutrinos their role in structure formation in our universe. The apparent enhancement of clustering, rather than the expected suppression, is present in a variety of CMB lensing observables. Furthermore, the conclusion is not altered by changes to the BAO and/or optical depth data even accounting for variations in the data that address potential systematic effects. It therefore seems likely that the trend will persist even with more data from near-term observations.

Even with current sensitivity, the preference for negative neutrino mass excludes $\sum m_\nu \geq 58$ meV at 3$\sigma$, thus making it a rather unlikely statistical fluctuation. Naturally, one would like to understand what might explain such a signal and how one could test these hypotheses. Of particular interest is whether this is just another hint of some breakdown in $\Lambda$CDM in the late-time expansion of the universe, rather than a change to the physics of neutrinos or structure formation directly. Here we showed how changes to the local expansion history are unlikely to fully explain the preference for negative neutrino mass without exacerbating existing tensions or creating new ones.

A variety of beyond the Standard Model (BSM) physics scenarios could ultimately explain these signals~\cite{Craig:2024tky}. Both $\sum m_\nu = 0$ and $\sum \tilde m_\nu < 0$ can arise within models that are otherwise weakly constrained by existing data. Given the likely persistence of this unusual signal, the most promising path to understanding the implications of current data may be to explore additional cosmological, astrophysical, and lab-based signals of these BSM models for hints of a underlying cause or for better constraints on the possibilities.

\paragraph{Acknowledgements}
We are grateful to Kim Berghaus, Emanuele Castorina, Eoin \'O Colg\'ain, Nathaniel Craig, Kyle Dawson, Simone Ferraro, Peter Graham, Lloyd Knox, Surjeet Rajendran, Uro\v s Seljak, Ben Wallisch, Deng Wang, Martin White, Matias Zaldarriaga for helpful discussions. DG is supported by the US~Department of Energy under grant~\mbox{DE-SC0009919}. JM is supported by the US~Department of Energy under grant~\mbox{DE-SC0010129}. Computational resources for this research were provided by SMU’s Center for Research Computing.  We acknowledge the use of \texttt{CAMB}~\cite{Lewis:1999bs}, \texttt{CLASS}~\cite{Blas:2011rf}, and the Python packages \texttt{Matplotlib}~\cite{Hunter:2007mat}, \texttt{NumPy}~\cite{Harris:2020xlr}, and~\texttt{SciPy}~\cite{Virtanen:2019joe}.

\newpage

\clearpage
\phantomsection
\addcontentsline{toc}{section}{References}
\bibliographystyle{utphys}
\bibliography{Refs}

\end{document}